\documentclass[prd,amsmath,amssymb,showpacs,superscriptaddress,nofootinbib,twocolumn]{revtex4-1}
\usepackage{amsmath}
\usepackage{graphicx}
\usepackage{epsfig,graphics,subfigure,psfrag,amsmath,amssymb}
\usepackage{lineno}
\usepackage{dcolumn}
\usepackage{bm}
\usepackage{overpic}
\usepackage{xspace}

\usepackage{rotating}
\usepackage{color}
\usepackage{multirow}
\usepackage{colortbl}
\usepackage{epstopdf}
\usepackage[colorlinks,linkcolor=blue,anchorcolor=blue,citecolor=blue]{hyperref}

\newcommand{\BR}{{\cal B}}

\newcommand{\pio}{\pi^{0}}

\begin{document}
\title{\boldmath Improved measurement of the branching fractions for $J/\psi\to\gamma\pio, \gamma\eta$ and $\gamma\eta^\prime$}

\author{M.~Ablikim$^{1}$, M.~N.~Achasov$^{5,b}$, P.~Adlarson$^{75}$, X.~C.~Ai$^{81}$, R.~Aliberti$^{36}$, A.~Amoroso$^{74A,74C}$, M.~R.~An$^{40}$, Q.~An$^{71,58}$, Y.~Bai$^{57}$, O.~Bakina$^{37}$
% ...
\\
 \vspace{0.2cm}
      (BESIII Collaboration)\\
      \vspace{0.2cm} {\it
$^{1}$ Institute of High Energy Physics, Beijing 100049, People's Republic of China\\
% ...
$^{5}$ Budker Institute of Nuclear Physics SB RAS (BINP), Novosibirsk 630090, Russia\\
% ...
$^{36}$ Johannes Gutenberg University of Mainz, Johann-Joachim-Becher-Weg 45, D-55099 Mainz, Germany\\
$^{37}$ Joint Institute for Nuclear Research, 141980 Dubna, Moscow region, Russia\\
% ...
$^{40}$ Liaoning Normal University, Dalian 116029, People's Republic of China\\
% ...
$^{57}$ Southeast University, Nanjing 211100, People's Republic of China\\
$^{58}$ State Key Laboratory of Particle Detection and Electronics, Beijing 100049, Hefei 230026, People's Republic of China\\
% ...
$^{71}$ University of Science and Technology of China, Hefei 230026, People's Republic of China\\
% ...
$^{74}$ University of Turin and INFN, (A)University of Turin, I-10125, Turin, Italy; (B)University of Eastern Piedmont, I-15121, Alessandria, Italy; (C)INFN, I-10125, Turin, Italy\\
$^{75}$ Uppsala University, Box 516, SE-75120 Uppsala, Sweden\\
% ...
$^{81}$ Zhengzhou University, Zhengzhou 450001, People's Republic of China\\
\vspace{0.2cm}
% ...
$^{b}$ Also at the Novosibirsk State University, Novosibirsk, 630090, Russia\\
% ...
}
}

\date{\today}


\begin{abstract}
Using a data sample of $(1.0087\pm 0.0044)\times 10^{10}$ $J/\psi$
events collected with the BESIII detector, the decays of
$J/\psi\to\gamma\pi^{0} (\eta, \eta^\prime)\to\gamma\gamma\gamma$ are
studied. Newly measured branching fractions are
$\mathcal{B}$$(J/\psi\to\gamma\pi^{0})$=$(3.34\pm 0.02\pm 0.09)\times
10^{-5}$, $\mathcal{B}$$(J/\psi\to\gamma\eta)$=$(1.096\pm 0.001\pm
0.019)\times 10^{-3}$ and
$\mathcal{B}$$(J/\psi\to\gamma\eta^\prime)$=$(5.40\pm 0.01\pm
0.11)\times 10^{-3}$, where the first uncertainties are statistical
and the second are systematic. These results are consistent with the
world average values within two standard deviations. The ratio of
partial widths
$\Gamma(J/\psi\to\gamma\eta^\prime)/\Gamma(J/\psi\to\gamma\eta)$ is
measured to be $4.93 \pm 0.13$. The singlet-octet pseudoscalar mixing
angle $\theta_P$ is determined to be $\theta_P = -(22.11 \pm
0.26)^\circ$ or $-(19.34 \pm 0.34)^\circ$ with two different
phenomenological models.
		
\end{abstract}

\maketitle

\section{Introduction}\label{sec:introduction}\vspace{-0.3cm}

Within the framework of Quantum Chromodynamics, the OZI-forbidden
radiative decays of $J/\psi\to\gamma\pio (\eta, \eta^\prime)$ are
expected to proceed predominantly via two virtual gluons which
subsequently convert to light hadrons, with the photon emitted from
the initial charm quarks~\cite{OZI}. These decays provide a clean
environment to study the conversion of gluons into hadrons and to test
various phenomenological
mechanisms~\cite{vdm1,vdm2,vdm3,mix4,mix5}. Of particular interest is
that the study of the radiative decays of $J/\psi\rightarrow\gamma\pio
(\eta, \eta^\prime)$ provide information on the quark composition of
the $\eta$ and $\eta^\prime$ mesons and the mixing between
them~\cite{uu,dd}.

Within flavor-SU(3) symmetry, the $\pio$, $\eta$ and $\eta^\prime$
mesons belong to the same pseudoscalar nonet. The physical states
$\eta$ and $\eta^\prime$ are commonly understood as mixtures of the
pure SU(3)-flavor octet
$[\eta_8=(u\bar{u}+d\bar{d}-2s\bar{s})/\sqrt{6}]$ and singlet
$[\eta_1=(u\bar{u}+d\bar{d}+s\bar{s})/\sqrt{3}]$ states,
\begin{equation}
\begin{split}
\eta=\eta_8\cos{\theta_P}-\eta_1\sin{\theta_P},\\%
\eta^\prime=\eta_8\sin{\theta_P}+\eta_1\cos{\theta_P},\label{Eq_mixing}
\end{split}	
\end{equation}
where $\theta_P$ is the pseudoscalar mixing angle
\cite{mixing}. The determination of this mixing parameter is
important because it allows us to understand the properties of
pseudoscalar mesons in terms of their underlying quark structure.

The radiative decays of $J/\psi\to\gamma\pio(\eta, \eta^\prime)$ have
been studied in many experiments~\cite{besII,CLEO,besIII2011}. The
most recent studies of $J/\psi\to\gamma\pi^0$, $J/\psi\to\gamma\eta$,
and $J/\psi\to\gamma\eta^\prime$ were reported by the BESIII
collaboration in
Refs.~\cite{BESIII2018axh,BESIII2021fos,BESIII2019gef}.

In this paper, using a data sample of $(1.0087\pm 0.0044)\times
10^{10}$ $J/\psi$ events~\cite{nJpsi0912} collected by the BESIII
detector, the branching fractions of $J/\psi\to\gamma\pi^0,
J/\psi\to\gamma\eta$ and $J/\psi\to\gamma\eta^\prime$ decays are
measured. The phenomenological
model-dependent mixing angles and the ratio of partial widths
$\Gamma(J/\psi\to\gamma\eta^\prime)/\Gamma(J/\psi\to\gamma\eta)$ are
also determined.

\section{Detector AND MONTE CARLO SIMULATION}\label{sec:dataset}
The BESIII detector~\cite{Ablikim:2009aa} records symmetric $e^+e^-$
collisions provided by the BEPCII storage
ring~\cite{Yu:IPAC2016-TUYA01} in the center-of-mass energy range from
2.0 to 4.95 GeV, with a peak luminosity of
$1\times10^{33}$~cm$^{-2}$s$^{-1}$ achieved at $\sqrt{s} = 3.77$
GeV. BESIII has collected large data samples in this energy
region~\cite{Ablikim:2019hff}. The cylindrical core of the BESIII
detector covers 93\% of the full solid angle and consists of a beam
pipe, a helium-based multilayer drift chamber~(MDC), a plastic
scintillator time-of-flight system~(TOF), and a CsI(Tl)
electromagnetic calorimeter~(EMC), which are all enclosed in a
superconducting solenoidal magnet providing a 1.0~T (0.9~T in 2012)
magnetic field. The solenoid is supported by an octagonal flux-return yoke
with resistive plate counter muon identification modules interleaved
with steel. The charged-particle momentum resolution at $1~{\rm
  GeV}/c$ is $0.5\%$, and the $dE/dx$ resolution is $6\%$ for
electrons from Bhabha scattering. The EMC measures photon energies
with a resolution of $2.5\%$ ($5\%$) at $1$~GeV in the barrel (end
cap) region. The time resolution in the TOF barrel region is 68~ps,
while that in the end cap region was 110~ps. The end cap TOF system was
upgraded in 2015 using multi-gap resistive plate chamber technology,
providing a time resolution of 60~ps~\cite{etof}.
	
Simulated samples produced with the {\sc geant4}-based~\cite{geant4}
simulation software~\cite{MCPackage} which includes the geometric
description~\cite{detvis} of the BESIII detector and the detector
response~\cite{BESIIIDetectorShape,BESIIIDetectorShapeB}, are used to
determine the detection efficiency and estimate backgrounds. The
inclusive Monte Carlo (MC) sample of $1.0\times 10^{10}$ simulated
inclusive $J/\psi$ events, used to estimate the background, includes
both the production of the $J/\psi$ resonance and the continuum
processes incorporated in {\sc kkmc}~\cite{kkmc}. The known decay
modes are modeled with {\sc evtgen}~\cite{evtgen} using branching
fractions taken from the Particle Data Group (PDG)~\cite{pdg}, and the
remaining unknown charmonium decays are modeled with {\sc
  lundcharm}~\cite{lundcharm}. To estimate the selection efficiency
and to optimize the selection criteria, 2.3 million MC signal events
for the $J/\psi\to\gamma\pio (\eta,
\eta^\prime)\to\gamma\gamma\gamma$ channels are generated, and the
decay modes are described with theoretical models that have been
validated in previous works~\cite{BESIII2018axh}. The polar angle of
the photon in the $J/\psi$ center-of-mass system is defined as
$\theta_{r}$, which follows $1+\cos^2\theta_{r}$ function. The
analysis is performed in the framework of the BESIII offline software
system~\cite{boss} which incorporates the detector calibration, event
reconstruction and data storage.

	\section{Event Selection and Background Analysis}
	
In this paper, the $\pio, \eta$ and $\eta^\prime$ mesons are all
reconstructed via their two-photon decays. Therefore, signal events
require at least three photons without any charged track in the
final states.
		
Photon candidates are reconstructed using clusters of energy deposits
in the EMC. The energy deposited in the nearby TOF is included to
improve the reconstruction efficiency and energy resolution. To exclude the background with small energy deposits in the EMC, the
photon candidates are required to have an energy greater than 80~MeV
in both the barrel region $(|\cos{\theta}| < 0.80)$ and endcap regions
$(0.86 < |\cos{\theta}| < 0.92)$.  A requirement on the EMC time
difference $\Delta T$ from the most energetic photon, $-500~ \rm ns <
\Delta T < 500$ ns, is used to suppress the electronic noise and
energy deposits unrelated to the event. Events with at least three
photon candidates are kept for further analysis.

A four-constraint ($4C$) kinematic fit imposing energy and momentum
conservation is performed under the hypothesis of $J/\psi\to\gamma\gamma\gamma$. If there is more than one
$\gamma\gamma\gamma$ combination, the one with the smallest
$\chi^2_{4C}$ from the kinematic fit is retained, and $\chi^2_{4C}<50$
is required. To reject backgrounds from the $e^+e^-\to\gamma\gamma$
process, similar 4C kinematic fit is performed on $\gamma\gamma$ combination, $\chi^2_{4C}$ is required to be less than the smallest
$\chi^2_{4C}(2\gamma)$. In the case of $J/\psi\to\gamma\pio$ decay, to
suppress the background from $J/\psi\to\gamma\pio\pio$ events,
in particular for events with missing low energy photons, $\chi^2_{4C}$ is required to be less than
the smallest $\chi^2_{4C}(4\gamma)$, where we require only four
photons to improve the efficiency.  This requirement is effective in
reducing 29\% of $J/\psi\to\gamma\pio\pio$ background events, while
removing no signal.
		
\begin{figure}[htbp]
		\begin{center}
\vskip -12pt
			\includegraphics[width=\columnwidth]{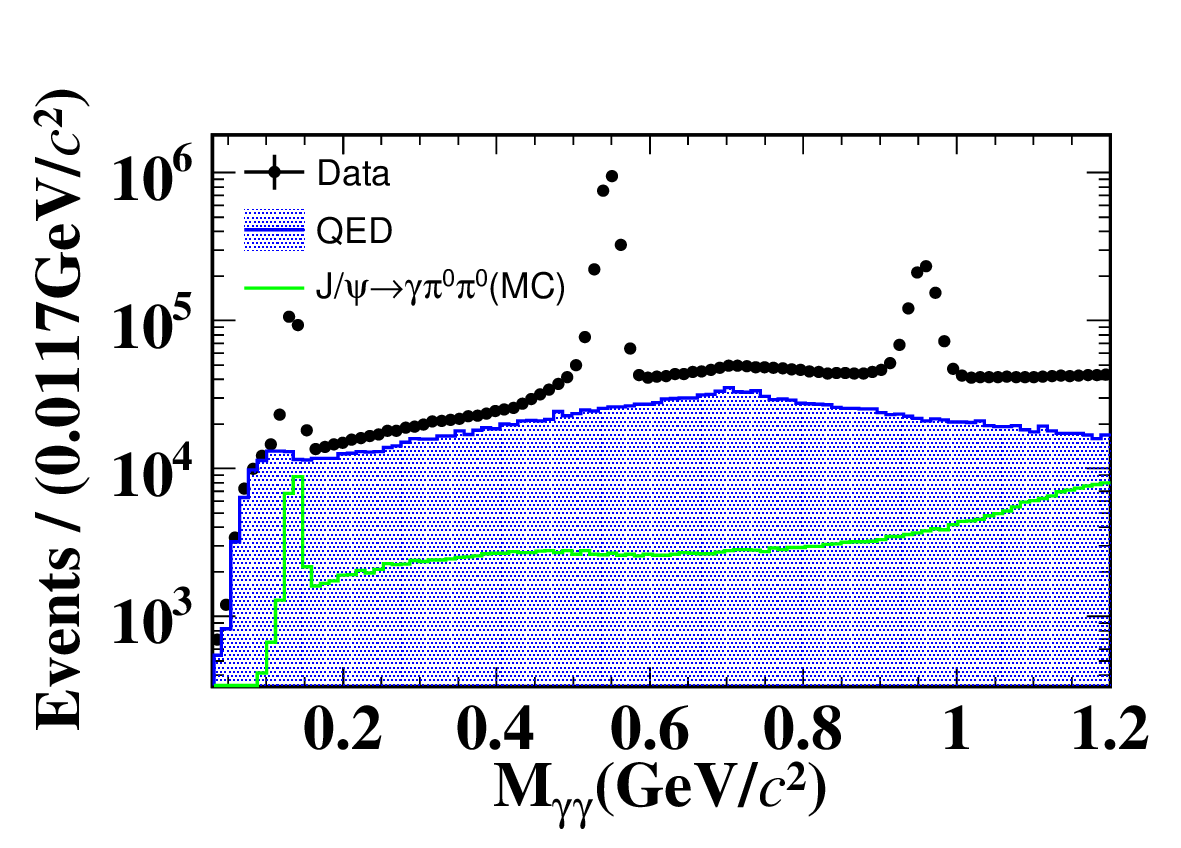}
\vskip -8pt
\caption{The $M_{\gamma\gamma}$ distributions for data (black dots
  with error bars), simulated background of $J/\psi\to\gamma\pio\pio$
  (green line) and QED background (blue filled area). The QED
  background is estimated from data taken at
  3.08~GeV.}\label{fig_QEDdata2pi0}
		\end{center}
		\end{figure}

After the above requirements, the distribution of the two-photon
invariant mass, $M_{\gamma\gamma}$, is shown in
Fig.~\ref{fig_QEDdata2pi0}, where the photon momenta from the 4C kinematic fit are used to calculate $M_{\gamma\gamma}$ and there are three entries per event, and   the $J/\psi\to\gamma\pio\pio$
channel is shown as a peaking background for $J/\psi\to\gamma\pio$
decay.
		
To estimate possible backgrounds, the inclusive MC sample and the 168.30~pb$^{-1}$ of data taken at
$\sqrt s=$ 3.08 GeV are used~\cite{lum12, lum1819}. From the inclusive
MC  sample, it is found that there is no peaking background for the $\eta$ and
$\eta^\prime$ peaks, but background events from
$J/\psi\to\gamma\pio\pio$ decay form a significant peak around the
nominal $\pi^0$ mass. To estimate its contribution, a dedicated MC
sample of $J/\psi\to\gamma\pio\pio$ events is produced in accordance
with the partial wave analysis results~\cite{gamma2pi0}. Using the
same selection criteria, and taking into account the number of
$J/\psi$ events and the branching fractions of
$J/\psi\to\gamma\pio\pio$~\cite{gamma2pi0} and
$\pio\to\gamma\gamma$~\cite{pdg}, the $M_{\gamma\gamma}$ distribution
is obtained and shown as a green line in Fig.~\ref{fig_QEDdata2pi0}.
The number of peaking background events in the $\pio$ signal region is $8208\pm134$, which is estimated by a fit to the $M_{\gamma\gamma}$ spectrum of the dedicated  MC
sample of $J/\psi\to\gamma\pio\pio$, where the $\pio$ signal is modeled with the
sum of Crystal Ball (CB)~\cite{CB} and Gaussian functions, while the
other non-peaking background is
described with a second-order Chebychev function.

For the background events directly from $e^+e^-$ annihilation, the
same analysis is performed on the data taken at the center-of-mass energy
of 3.08 GeV. The selected events are normalized to the $J/\psi$ data
sample, after taking into account the luminosities and
energy-dependent cross sections of the quantum electrodynamics (QED)
processes, with a factor $f$
       \begin{linenomath*}
       \begin{equation}
		    f\equiv \frac{N_{3.097}}{N_{3.080}} = \frac{{\cal
                      L}_{3.097}}{{\cal L}_{3.080}} \cdot
                    \frac{\sigma_{3.097}}{\sigma_{3.080}} \cdot
                    \frac{\varepsilon_{3.097}}{\varepsilon_{3.080}},\label{Eq_QED}
		\end{equation}
        \end{linenomath*}
where $N$, ${\cal L}$, $\sigma$ and $\varepsilon$ refer to signal yields,
integrated luminosities of data samples, cross sections and detection
efficiencies at the two center-of-mass energies, respectively. The
details on the cross sections can be found in Ref.~\cite{QED}.

The normalized $M_{\gamma\gamma}$ spectrum from $e^+e^-$ annihilation,
obtained from the data at the center-of-mass energy of 3.080 GeV, is
illustrated as the filled area in Fig.~\ref{fig_QEDdata2pi0}.  Due to
identical event topology, these background events are
indistinguishable from signal events.  No obvious peaking contribution
in the $\pi^0$, $\eta$ and $\eta^\prime$ signal regions is
observed. Interference with the resonance amplitude is expected to be
small and is ignored.

\begin{figure}[htbp]
		\begin{center}
			\subfigure{\includegraphics[width=\columnwidth]{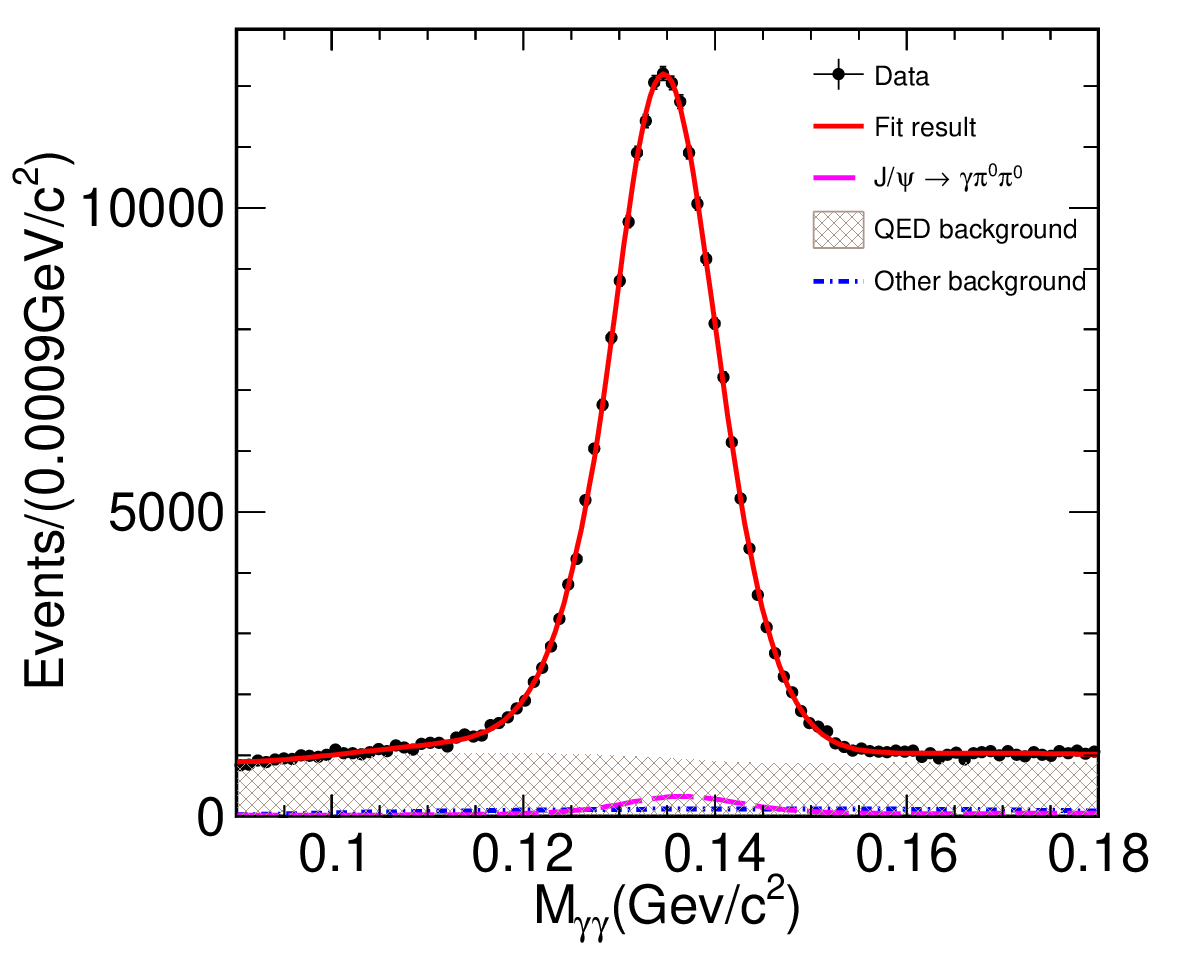}\label{fig_pi0_fitdata}}
			\put(-180,168){(a)}\\
                \vskip -5pt

        \subfigure{\includegraphics[width=\columnwidth]{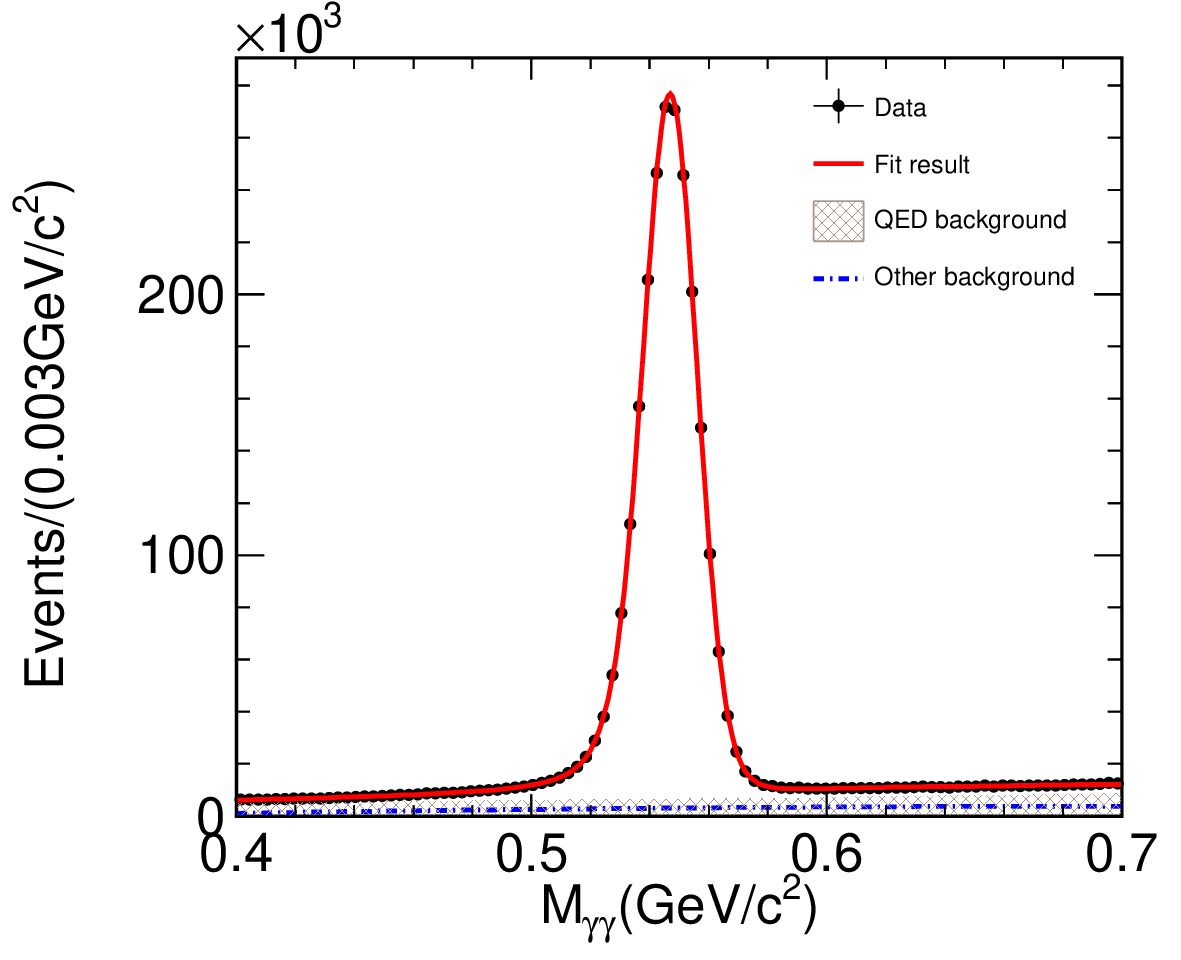}\label{fig_eta_fitdata_draft}}
			\put(-180,165){(b)}\\
              \vskip -3pt
			\subfigure{\includegraphics[width=\columnwidth]{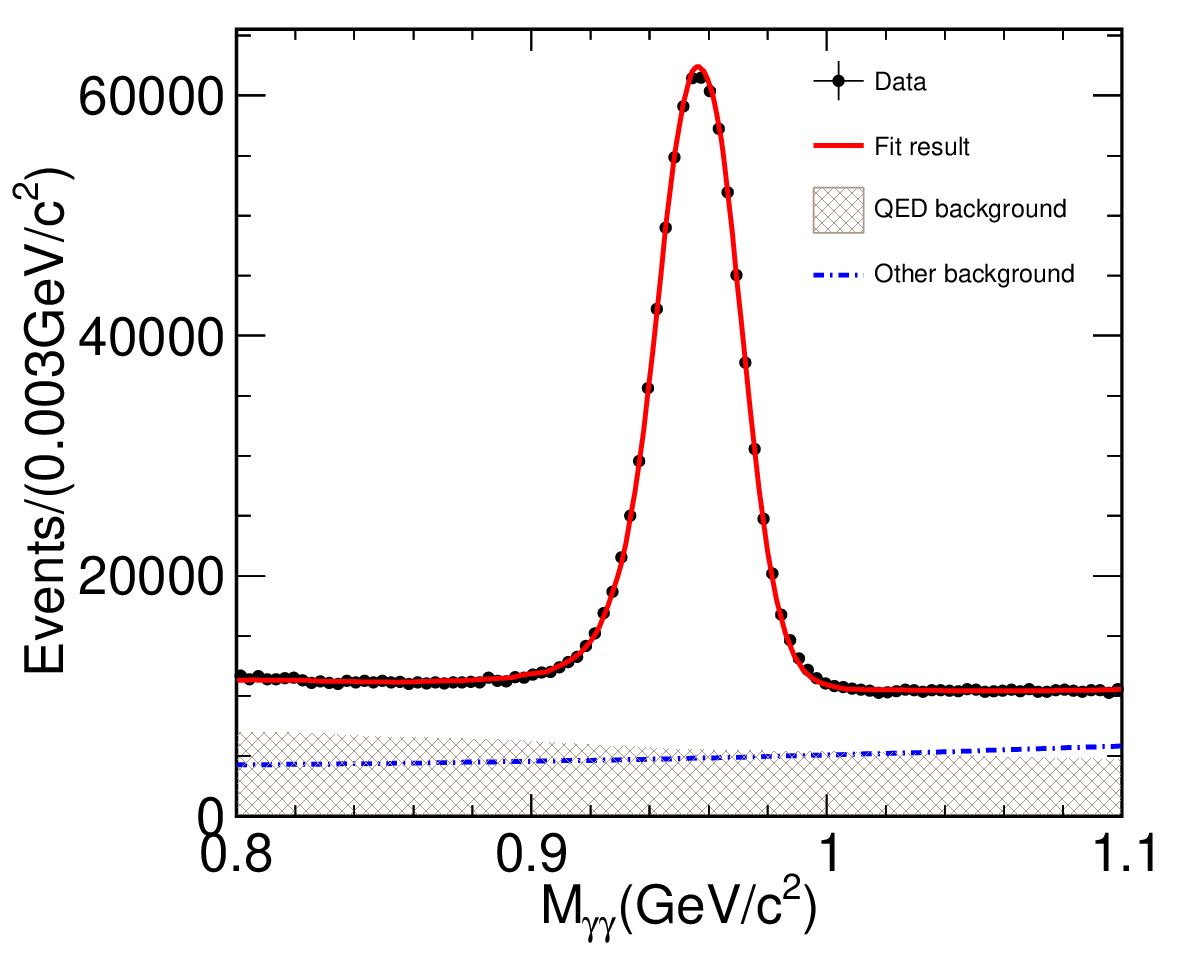}\label{fig_etap_fitdata_draft}}
			\put(-180,168){(c)}\\
\vskip -10pt
\caption{The fit results of the $M_{\gamma\gamma}$ spectra for the decays of (a)
  $J/\psi\to\gamma\pio\to 3\gamma$, (b) $J/\psi\to\gamma\eta\to
  3\gamma$ and (c) $J/\psi\to\gamma\eta^\prime\to 3\gamma$. Dots
  with error bars represent data, the brown filled area represents QED
  background, the pink and blue dot-dashed lines describe a
  contribution from $J/\psi\to\gamma\pio\pio$ and other background
  contributions, respectively.}\label{fig_fit}
		\end{center}
		\end{figure}

The signal yields are obtained from unbinned maximum likelihood fits
to the $M_{\gamma\gamma}$ distributions in the $\pi^0$, $\eta$ and
$\eta^\prime$ mass regions, as shown in Fig.~\ref{fig_fit}.  For each
fit, the total probability density function consists of a signal and
background contributions. To properly describe the data, the
$\pi^0$($\eta$) signal component is modeled using a sum of CB and
Gaussian functions, while the $\eta^\prime$ peak is modeled using the
MC-simulated shape convoluted with a Gaussian function to account for
the mass resolution difference between data and MC simulation.  The
background components are: (i) the MC-simulated
$J/\psi\rightarrow\gamma\pi^0\pi^0$ background; (ii) the QED
background derived from the data taken at the center-of-mass of 3.08
GeV; (iii) the remaining background events are parameterized with a
second-order Chebychev function.  The fitted signal yields and the
detection efficiencies are summarized in Table~\ref{table_result}.

		\begin{table*}[htbp]
		\begin{center}
\caption{The branching fractions and comparison with previous
  results. The first uncertainty is statistical and the second is
  systematic. $P$ represents the pseudoscalar meson $\pi^0 (\eta,
  \eta^\prime)$.}\label{table_result}
\setlength{\tabcolsep}{0.4cm}{\begin{tabular}{lccccc}\hline\hline & &
    & \multicolumn{3}{c}{$\mathcal{B}(J/\psi\to\gamma P)~(\times
      10^{-4})$}\\\cline{4-6} \multicolumn{1}{c}{$\gamma P$} &
    \multicolumn{1}{c}{$N^{\rm{obs}}_{J/\psi\to\gamma P}$} &
    \multicolumn{1}{c}{$\varepsilon$(\%)} & \multicolumn{1}{c}{This
      work} & \multicolumn{1}{c}{BESIII} &
    \multicolumn{1}{c}{PDG~\cite{pdg}}\\\hline $\gamma\pio$ &
    175893$\pm$839 & 52.90$\pm$0.03 & 0.334$\pm$0.002$\pm$0.009 &
    0.361$\pm$0.012$\pm$0.016~\cite{BESIII2018axh} & 0.356$\pm$0.017\\ $\gamma\eta$ &
    2209063$\pm$2592 & 50.78$\pm$0.03 & 10.96$\pm$0.01$\pm$0.19 &
    $10.67\pm0.05\pm0.23$~\cite{BESIII2021fos} & 10.85$\pm$0.18\\ $\gamma\eta^\prime$ &
    638206$\pm$1061 & 50.77$\pm$0.03 & 54.0$\pm$0.1$\pm$1.1 &
    $52.7\pm0.3\pm0.5$~\cite{BESIII2019gef} & 52.5$\pm$0.7\\ \hline\hline
			\end{tabular}}
		\end{center}
		\end{table*}

\section{Systematic Uncertainties}\label{sec:sysU}
The systematic uncertainties of the branching fraction measurements
originate mainly from the photon detection and kinematic fit
efficiencies. Additional uncertainties associated with the fit method,
signal shape, background of $J/\psi\to\gamma\pio\pio$, QED background,
branching fractions of $\pio(\eta ,\eta^\prime)\to\gamma\gamma$ and
number of $J/\psi$ events are also considered.

	    \begin{itemize}
	    \item {\bf Fit method}
	
The uncertainty associated with the fit method comes from two sources:
the fit range and background shape.  The uncertainty arising from the
fit range is estimated by adjusting the fit range by $\pm5$ MeV. The
maximum differences in the branching fractions with respect to the
baseline values are taken as the systematic uncertainties.  To
estimate the uncertainty associated with the background shape,
alternative fits are performed using first-order or third-order
Chebychev functions. The maximum changes of the fitted signal yields
are taken as the systematic uncertainties, which are summarized in
Table~\ref{table_sysSumInc}.

	    \end{itemize}

     \begin{itemize}
	    \item {\bf Signal shape}

To evaluate the systematic uncertainty arising from the signal shape,
alternative fits are performed using the MC-simulated shape to
describe the signal component of $J/\psi\to\gamma \pio (\eta)$ decays
and a sum of CB and Gaussian functions for
$J/\psi\to\gamma\eta^\prime$ decay. The differences between the
results with the baseline and alternative methods are taken as the
corresponding systematic uncertainties.

     \end{itemize}

\begin{itemize}
\item {\bf \boldmath Background of $J/\psi\to\gamma\pio\pio$} The background
  yield of $J/\psi\to\gamma\pio\pio$ is fixed in the fit according to
  the branching fraction from Ref.~\cite{gamma2pi0}. Alternative fits
  are performed varying the input branching fraction by one standard
  deviation. The biggest difference of the fitted signal yield with
  respect to the baseline result, 0.21\%, is taken as a systematic
  uncertainty.

	    \end{itemize}	

	    \begin{itemize}
\item {\bf QED background }
	
The QED background is estimated with Eq.~(\ref{Eq_QED}) using the data
taken at the center-of-mass energy of 3.080 GeV. Alternative fits are
performed varying ${\cal L}_{3.080}$ in Eq.~(\ref{Eq_QED}) by one
standard deviation. The differences from the nominal fits are taken as
the systematic uncertainties.
		
	    \end{itemize}
	
	    \begin{itemize}
\item {\bf Photon detection}
	
The systematic uncertainty of the photon detection efficiency is
studied using a control sample of $e^+e^-\to\gamma\mu^+\mu^-$
events. The four-momentum of the initial-state-radiation photon is
predicted by the $\mu^+\mu^-$ pair. The photon detection efficiency is
defined as the fraction of reconstructed photons with four-momentum
matching in the EMC. The systematic uncertainty is defined as the
relative difference in efficiency between data and MC simulation,
which is estimated using a re-weighting
technique~\cite{reweighting}. The uncertainties of the photon
detection for the $J/\psi\to\gamma\pio$, $J/\psi\to\gamma\eta$ and
$J/\psi\to\gamma\eta^\prime$ decays are 0.57\%, 0.51\% and 0.47\%,
respectively.
	    \end{itemize}
	
	    \begin{itemize}
	    \item {\bf Kinematic fit}

The uncertainty of the kinematic fit mainly comes from the
inconsistency of the photon resolution between data and MC
simulation. We adjust the energy resolution in the reconstructed
photon error matrix so that the MC simulation provides a good
description of data~\cite{uncertaitykinematicfit}. The fits are redone,
and the changes
with respect to the nominal results are taken as the systematic
uncertainties.
		
	    \end{itemize}
	
	    \begin{itemize}
\item {\bf Intermediate branching fractions}

The branching fractions of intermediate decays are taken from the
PDG~\cite{pdg}. The uncertainties of the branching fractions of
$\pi^0\to \gamma\gamma$, $\eta\to \gamma\gamma$ and $\eta^\prime \to
\gamma\gamma$ are 0.03\%, 0.46\%, and 1.43\%, respectively.
	    \end{itemize}
	
	    \begin{itemize}
\item {\bf \boldmath Number of $J/\psi$ events}
	
The number of $J/\psi$ events is determined to be $(1.0087\pm
0.0044)\times10^{10}$ using inclusive hadronic $J/\psi$
decays~\cite{nJpsi0912}. Its uncertainty, 0.44\%, is taken as a
systematic uncertainty.
	    \end{itemize}

The systematic uncertainties are summarized in
Table~\ref{table_sysSumInc}. The total systematic uncertainty is given
by the square root of the quadratic sum of the individual
contributions.
	
	    \begin{table}[htbp]
		\begin{center}
			\caption{Systematic sources and corresponding contributions (\%).}\label{table_sysSumInc}
			\begin{tabular}{lccc}\hline\hline
				Source\color{white}{\Large{1}} &~ $\gamma\pio$ &~ $\gamma\eta$ &~ $\gamma\eta^\prime$\\\hline
                Fit range  &~0.9 &~0.8 &~1.08\\
                Background Shape &~1.13 &~1.07 &~0.39\\
                Signal shape &~ 1.99 &~ 0.11 &~ 0.49\\
                Background of $J/\psi\to\gamma\pio\pio$ &~ 0.21 &~ -- &~ --\\
		QED background &~ 1.08 &~ 0.15 &~ 0.01\\
		Photon detection &~ 0.57 &~ 0.51 &~ 0.47\\
		$4C$ kinematic fit &~ 0.41 &~ 0.80 &~ 0.53\\
		Intermediate branching fractions &~ 0.03 &~ 0.46 &~ 1.43\\
		$N_{J/\psi}$ &~ 0.44 &~ 0.44 &~ 0.44\\\hline
		Total &~ 2.82 &~ 1.77 &~ 2.07\\
				\hline\hline
			\end{tabular}
		\end{center}
		\end{table}

	\section{Results}
	    The branching fraction of $J/\psi\to\gamma P$ is calculated as
                \begin{linenomath*}
			\begin{equation}
		  \mathcal{B}(J/\psi\to\gamma P) = \frac{N_{\rm obs}}{N_{J/\psi}\times \BR(P\to\gamma\gamma)\times \varepsilon},\label{Eq_BR}
		    \end{equation}
                \end{linenomath*}
where $P$ represents the pseudoscalar meson $\pi^0 (\eta,
\eta^\prime)$, $N_{\rm obs}$ is the number of the observed signal
events, $N_{J/\psi}$ is the number of $J/\psi$ events,
$\varepsilon$ is the detection efficiency and $\BR(P\to\gamma\gamma)$
is the branching fraction of $P\to\gamma\gamma$~\cite{pdg}.
		
The measured branching fractions using Eq.~(\ref{Eq_BR}) are
summarized in Table~\ref{table_result}. They are in
agreement with the previous BESIII
results~\cite{BESIII2021fos,BESIII2019gef,BESIII2018axh} and the world
average values~\cite{pdg}. The results allow the study of
pseudoscalar mixing in Eq.~\ref{Eq_mixing}.  Under the assumption of
exact SU(3) flavor symmetry, the mixing angle can be determined via
                \begin{linenomath*}
		    \begin{equation}
		    R=\frac{\Gamma(J/\psi\to\gamma\eta^\prime)}{\Gamma(J/\psi\to\gamma\eta)}=\left(\frac{p_{\eta^\prime}}{p_\eta}\right)^3\cdot\cot^2{\theta_P}\label{Eq_R},
		    \end{equation}
                \end{linenomath*}
where $p_\eta$ and $p_{\eta^\prime}$ are the momenta of $\eta$ and
$\eta^\prime$ in the $J/\psi$ rest frame, respectively. The systematic
uncertainties caused by the number of $J/\psi$ events and photon
detection efficiency can be eliminated in the ratio
calculation. According to the theoretical calculation~\cite{dd},
the $\theta_P$ value is negative. Finally, we obtain $\theta_P =
-(22.11 \pm 0.26)^\circ$, where the uncertainty includes both statistical and systematic uncertainties.
		
The $\eta-\eta^\prime$ mixing approach has been generalized to also
include other states, such as $\pio$ meson. The decay width of
$J/\psi\to\gamma P$ is given by~\cite{mixing angle}
                \begin{linenomath*}
		    \begin{equation}
		    \Gamma(J/\psi\to\gamma P)=\frac{1}{3}\frac{g^2_{\gamma P}}{4\pi}\frac{p^3}{m^2_{J/\psi}},
		    \end{equation}
                \end{linenomath*}
where $g^2_{\gamma P}$ is the coupling constant defined by the
amplitude, as shown in Table~\ref{table_mixing}, and $p$ is the
magnitude of momentum of $P$. In this table, $d$ and $f$ coefficients
are free parameters, derived from a joint set of decay width
equations, which cancel in the calculation, and the
details can be found in the appendix. The $X_{\eta(\eta^\prime)}$
and $Y_{\eta(\eta^\prime)}$ are related to the following:
    \begin{linenomath*}
      \begin{equation}
      \begin{split}
          &X_\eta=Y_{\eta^\prime}=\cos\theta_P/\sqrt{3}-\sqrt{\frac{2}{3}}\sin\theta_P,\\
          &Y_\eta=-X_{\eta^\prime}=-\sqrt{\frac{2}{3}}\cos\theta_P-\sin\theta_P/\sqrt{3},\\%
                \end{split}
		    \end{equation}
                \end{linenomath*}

The $\theta_P$ value is determined to be $-(19.34 \pm 0.34)^\circ$, where the uncertainty includes both statistical and systematic uncertainties.
		
   \begin{table}[htbp]
		\begin{center}
			\caption{ Amplitudes for the decays of $J/\psi\to\gamma P$.}\label{table_mixing}
			\begin{tabular}{c|c}\hline\hline
				Decay mode & Amplitude\\\hline
				$\gamma\pio$ & $\frac{f}{\sqrt{6}}$\\
				
				$\gamma\eta$ & $\frac{2}{\sqrt{6}}( d+\frac{f}{3})X_\eta+\frac{1}{\sqrt{3}}( d-\frac{f}{3})Y_\eta$\\
				
				$\gamma\eta^\prime$ & $\frac{2}{\sqrt{6}}( d+\frac{f}{3})X_{\eta^\prime}+\frac{1}{\sqrt{3}}( d-\frac{f}{3})Y_{\eta^\prime}$\\
				\hline\hline
			\end{tabular}
		\end{center}
		\end{table}		
			
\section{Summary}\label{sec:summary}
Based on $(1.0087\pm 0.0044)\times 10^{10}$ $J/\psi$ events collected
with the BESIII detector, a study of $J/\psi\to\gamma\pio (\eta,
\eta^\prime)$ decays is performed. The branching fractions of
$J/\psi\to\gamma \pio$, $J/\psi\to\gamma\eta$ and
$J/\psi\to\gamma\eta^\prime$ are measured to be $(3.34\pm 0.02\pm
0.09)\times 10^{-5}$, $(1.096\pm 0.001\pm 0.019)\times 10^{-3}$ and
$(5.40\pm 0.01\pm 0.11)\times 10^{-3}$, respectively, which are
consistent with the previous
measurements~\cite{besII,CLEO,BESIII2021fos,BESIII2019gef,BESIII2018axh}
within two standard deviations. As shown in Table~\ref{table_result},
the result for $B(J/\psi\to\gamma \pio)$ has much better precision compared with the previous BESIII result, while $B(J/\psi\to\gamma\eta)$ also has better precision.
 With two phenomenological
models~\cite{R}, the singlet-octet pseudoscalar mixing angles of
$\theta_P$ are determined to be $\theta_P = -(22.11 \pm 0.26)^\circ$
and $-(19.34 \pm 0.34)^\circ$, respectively, which are consistent with
the theoretical calculation of Ref. \cite{dd}.

\begin{acknowledgements}
The BESIII Collaboration thanks the staff of BEPCII and the IHEP computing center for their strong support. This work is supported in part by National Key R$\&$D Program of China under Contracts Nos. 2020YFA0406300, 2020YFA0406400; National Natural Science Foundation of China (NSFC) under Contracts Nos.  12105132, 11905092, 12225509,  11635010, 11735014, 11835012, 11935015, 11935016, 11935018, 11961141012, 12022510, 12025502, 12035009, 12035013, 12061131003, 12192260, 12192261, 12192262, 12192263, 12192264, 12192265, 12221005, 12235017, 11705078; the Chinese Academy of Sciences (CAS) Large-Scale Scientific Facility Program; the CAS Center for Excellence in Particle Physics (CCEPP); Joint Large-Scale Scientific Facility Funds of the NSFC and CAS under Contract No. U1832207; CAS Key Research Program of Frontier Sciences under Contracts Nos. QYZDJ-SSW-SLH003, QYZDJ-SSW-SLH040; 100 Talents Program of CAS; The Institute of Nuclear and Particle Physics (INPAC) and Shanghai Key Laboratory for Particle Physics and Cosmology; ERC under Contract No. 758462; European Union's Horizon 2020 research and innovation programme under Marie Sklodowska-Curie grant agreement under Contract No. 894790; German Research Foundation DFG under Contracts Nos. 443159800, 455635585, Collaborative Research Center CRC 1044, FOR5327, GRK 2149; Istituto Nazionale di Fisica Nucleare, Italy; Ministry of Development of Turkey under Contract No. DPT2006K-120470; National Research Foundation of Korea under Contract No. NRF-2022R1A2C1092335; National Science and Technology fund of Mongolia; National Science Research and Innovation Fund (NSRF) via the Program Management Unit for Human Resources \& Institutional Development, Research and Innovation of Thailand under Contract No. B16F640076; Polish National Science Centre under Contract No. 2019/35/O/ST2/02907; The Swedish Research Council; U. S. Department of Energy under Contract No. DE-FG02-05ER41374; China Postdoctoral Science Foundation under Contracts Nos. 2021M693181; The PhD Start-up Foundation of Liaoning Province under Contracts No. 2019-BS-113; Education Department of Liaoning Province Scientific research Foundation under Contracts No. LQN201902; Foundation of Innovation team 2020, Liaoning Province; Opening Foundation of Songshan Lake Materials Laboratory, Grants No.2021SLABFK04.
\end{acknowledgements}

\section*{\boldmath Appendix: Calculation of $\theta_P$}

The decay width of $J/\psi\to\gamma P$ is given by the Eq. (5).
One can obtain that:

		    \begin{equation}
		    \frac{\Gamma(J/\psi\to\gamma \eta)}{\Gamma(J/\psi\to\gamma \pi^0)}=\frac{g^2_{\gamma \eta}}{g^2_{\gamma \pi^0}}\frac{p_{\eta}^3}{p_{\pi^0}^3} = \frac{ \mathcal{B}(J/\psi\to\gamma\eta)}{ \mathcal{B}(J/\psi\to\gamma\pi^0)}
		    \end{equation}

		    \begin{equation}
		    \frac{\Gamma(J/\psi\to\gamma \eta')}{\Gamma(J/\psi\to\gamma \pi^0)}=\frac{g^2_{\gamma \eta'}}{g^2_{\gamma \pi^0}}\frac{p_{\eta'}^3}{p_{\pi^0}^3} = \frac{ \mathcal{B}(J/\psi\to\gamma\eta')}{ \mathcal{B}(J/\psi\to\gamma\pi^0)}
		    \end{equation}

\noindent where $g^2_{\gamma P}$ is the coupling constant defined by the amplitude, as shown in Table~\ref{table_mixing}, and $p$ is the magnitude of momentum of $P$. Then we have:

  \begin{widetext}
        \begin{flalign}
      \begin{split}
      		\sqrt{\frac{ \mathcal{B}(J/\psi\to\gamma\eta)}{\mathcal{B}(J/\psi\to\gamma\pi^0)}\frac{p_{\pi^0}^3}{p_{\eta}^3} }
      &= \frac{g_{\gamma \eta}}{g_{\gamma \pi^0}} \\ 
      &= \frac{\frac{2}{\sqrt{6}}( d+\frac{f}{3})X_\eta+\frac{1}{\sqrt{3}}( d-\frac{f}{3})Y_\eta}{\frac{f}{\sqrt{6}}}\\ 
      &=\frac{\frac{2}{\sqrt{6}}( d+\frac{f}{3})(\cos\theta_P/\sqrt{3}-\sqrt{\frac{2}{3}}\sin\theta_P)+\frac{1}{\sqrt{3}}( d-\frac{f}{3})(-\sqrt{\frac{2}{3}}\cos\theta_P-\sin\theta_P/\sqrt{3})}{\frac{f}{\sqrt{6}}} \\
      &= \frac{-9\sqrt{2}d\sin\theta_P+4f\cos\theta_P-\sqrt{2}f\sin\theta_P}{3\sqrt{3}f} \\  
          \end{split}&
            \end{flalign}

  \begin{flalign}
    &\  3\sqrt{3}f\cos\theta_P \sqrt{\frac{ \mathcal{B}(J/\psi\to\gamma\eta)}{ \mathcal{B}(J/\psi\to\gamma\pi^0)}\frac{p_{\pi^0}^3}{p_{\eta}^3}}  =
       -9\sqrt{2}d\sin\theta_P\cos\theta_P+4f\cos^{2}\theta_P-\sqrt{2}f\sin\theta_P\cos\theta_P &
      \end{flalign}

      \begin{flalign}
      \begin{split}
      \sqrt{\frac{ \mathcal{B}(J/\psi\to\gamma\eta')}{ \mathcal{B}(J/\psi\to\gamma\pi^0)}\frac{p_{\pi^0}^3}{p_{\eta'}^3}} &= \frac{g_{\gamma \eta'}}{g_{\gamma \pi^0}} \\
      &=\frac{\frac{2}{\sqrt{6}}( d+\frac{f}{3})X_\eta'+\frac{1}{\sqrt{3}}( d-\frac{f}{3})Y_\eta'}{\frac{f}{\sqrt{6}}}\\
    &=\frac{\frac{2}{\sqrt{6}}( d+\frac{f}{3})(\sin\theta_P/\sqrt{3}+\sqrt{\frac{2}{3}}\cos\theta_P)+\frac{1}{\sqrt{3}}( d-\frac{f}{3})(-\sqrt{\frac{2}{3}}\sin\theta_P+\cos\theta_P/\sqrt{3})}{\frac{f}{\sqrt{6}}}\\
   &= \frac{9\sqrt{2}d\cos\theta_P+4f\sin\theta_P+\sqrt{2}f\cos\theta_P}{3\sqrt{3}f}\\
        \end{split}&
        \end{flalign}

      \begin{flalign}
     &\ 3\sqrt{3}f\sin\theta_P \sqrt{\frac{ \mathcal{B}(J/\psi\to\gamma\eta')}{ \mathcal{B}(J/\psi\to\gamma\pi^0)}\frac{p_{\pi^0}^3}{p_{\eta'}^3}}  = 9\sqrt{2}d\sin\theta_P\cos\theta_P+4f\sin^{2}\theta_P+\sqrt{2}f\sin\theta_P\cos\theta_P &
            \end{flalign}

With Eqs. (10) and (12), one can obtain that:

\begin{equation}
      \begin{split}
      3\sqrt{3}f\cos\theta_P \sqrt{\frac{ \mathcal{B}(J/\psi\to\gamma\eta)}{ \mathcal{B}(J/\psi\to\gamma\pi^0)}\frac{p_{\pi^0}^3}{p_{\eta}^3}} &+  3\sqrt{3}f\sin\theta_P \sqrt{\frac{ \mathcal{B}(J/\psi\to\gamma\eta')}{ \mathcal{B}(J/\psi\to\gamma\pi^0)}\frac{p_{\pi^0}^3}{p_{\eta'}^3}} = 4f(\sin^{2}\theta_P+\cos^{2}\theta_P)
   \end{split}
            \end{equation}

   \begin{equation}
      \begin{split}
      \frac{3\sqrt{3}}{4}\cos\theta_P \sqrt{\frac{ \mathcal{B}(J/\psi\to\gamma\eta)}{ \mathcal{B}(J/\psi\to\gamma\pi^0)}\frac{p_{\pi^0}^3}{p_{\eta}^3}} &+  \frac{3\sqrt{3}}{4}\sin\theta_P \sqrt{\frac{ \mathcal{B}(J/\psi\to\gamma\eta')}{ \mathcal{B}(J/\psi\to\gamma\pi^0)}\frac{p_{\pi^0}^3}{p_{\eta'}^3}} = 1
      \end{split}
            \end{equation}

The Eq. $(14)$ can be written as:
\begin{equation}
      \begin{split}
      X\cos\theta_P+Y\sin\theta_P = 1
   \end{split}
            \end{equation}

Where  $X=\frac{3\sqrt{3}}{4}\sqrt{\frac{ \mathcal{B}(J/\psi\to\gamma\eta)}{ \mathcal{B}(J/\psi\to\gamma\pi^0)}\frac{p_{\pi^0}^3}{p_{\eta}^3}}$,
            $Y = \frac{3\sqrt{3}}{4}\sqrt{\frac{ \mathcal{B}(J/\psi\to\gamma\eta')}{ \mathcal{B}(J/\psi\to\gamma\pi^0)}\frac{p_{\pi^0}^3}{p_{\eta'}^3}}$. Then:

         \begin{equation}
      \begin{split}
      &X\sin\theta_P-Y\cos\theta_P \\&= \pm\sqrt{ X^2\sin^2\theta_P+Y^2\cos^2\theta_P-2XY\sin\theta_P\cos\theta_P}\\
      &=\pm\sqrt{ X^2\sin^2\theta_P+Y^2\cos^2\theta_P-[(X\cos\theta_P+Y\sin\theta_P)^2-(X^2\cos^2\theta_P+Y^2\sin^2\theta_P)]}\\
      &=\pm\sqrt{X^2(\sin^2\theta_P+\cos^2\theta_P)+Y^2(\sin^2\theta_P+\cos^2\theta_P)-1}\\
      &=\pm\sqrt{X^2+Y^2-1}
   \end{split}
            \end{equation}

 According to Eqs. (15) and (16), one can obtain that:
      \begin{align}
     \left\{
     \begin{aligned}
     &X\cos\theta_P+Y\sin\theta_P = 1\\
     &X\sin\theta_P-Y\cos\theta_P = \pm\sqrt{X^2+Y^2-1}
     \end{aligned}
     \right.
   \end{align}
    \begin{align}
     \left\{
     \begin{aligned}
     &Y(X\cos\theta_P+Y\sin\theta_P) = Y\\
     &X(X\sin\theta_P-Y\cos\theta_P) = \pm X\sqrt{X^2+Y^2-1}
     \end{aligned}
     \right.
   \end{align}

   \begin{equation}
      (X^2+Y^2)\sin\theta_P = Y\pm X\sqrt{X^2+Y^2-1}
            \end{equation}
   \begin{equation}
      \sin\theta_P = \frac{Y\pm X\sqrt{X^2+Y^2-1}}{X^2+Y^2}
            \end{equation}

 According to the theoretical calculation of
Ref.~\cite{dd}, the value of $\theta_P$ is negative.
 \begin{equation}
      \theta_P = \arcsin(\frac{Y - X\sqrt{X^2+Y^2-1}}{X^2+Y^2})
            \end{equation}
 \end{widetext}

\end{document}